\documentclass[aip,jap,amsmath,amssymb,reprint,onecolumn]{revtex4-1}

\usepackage{graphicx}
\usepackage{dcolumn}
\usepackage{bm}
\usepackage{epsfig}

\newlength{\epigraphwidth}
\begin{document}


\title{Challenges in nucleosynthesis of trans-iron elements}

\author{T. Rauscher}%
\affiliation{%
Centre for Astrophysics Research, School of Physics, Astronomy and Mathematics, Hatfield AL10 9AB, United Kingdom}%
\affiliation{%
Department of Physics, University of Basel, CH-4056 Basel, Switzerland}%


\begin{abstract}
Nucleosynthesis beyond Fe poses additional challenges not encountered when studying astrophysical processes involving light nuclei. Astrophysical sites and conditions are not well known for some of the processes involved. On the nuclear physics side, different approaches are required, both in theory and experiment. The main differences and most important considerations are presented for a selection of nucleosynthesis processes and reactions, specifically the $s$-, $r$-, $\gamma$-, and $\nu p$-processes. Among the discussed issues are uncertainties in sites and production conditions, the difference between laboratory and stellar rates, reaction mechanisms, important transitions, thermal population of excited states, and uncertainty estimates for stellar rates. The utility and limitations of indirect experimental approaches are also addressed. The presentation should not be viewed as confining the discussed problems to the specific processes. The intention is to generally introduce the concepts and possible pitfalls along with some examples. Similar problems may apply to further astrophysical processes involving nuclei from the Fe region upward and/or at high plasma temperatures. The framework and strategies presented here are intended to aid the conception of future experimental and theoretical approaches.
\end{abstract}


\maketitle

\hfill \begin{minipage}{1.2\epigraphwidth}
{\vspace{1cm}
\flushright \footnotesize \it
The only way of discovering the limits of the possible\\
is to venture a little way past them into the impossible.\\
\hfill \textsc{Arthur C. Clarke}
}
\end{minipage}

\section{Introduction}
\label{sec:intro}

Elements up to the Iron peak in the solar abundance distribution can be made in hydrostatic stellar burning processes, whereas heavier elements require extremer conditions, such as explosions resulting from thermonuclear burning or rapid ejection and decompression of gravitationally condensed and heated material. The former are connected to He-shell flashes in stars with less than 8 solar masses and to type Ia supernovae, the latter are realized in core-collapse supernovae and neutron star mergers. Although nuclear processes under explosive conditions -- depending on the specific phenomenon -- may also produce nuclei around and below the Fe-Ni region, heavier nuclides cannot be produced in conditions encountered in hydrostatic burning. Due to the different temperature and density ranges of explosive burning with respect to hydrostatic burning and due to the fact that heavier nuclei are involved, also different aspects of the nuclear processes have to be studied and the behavior of reaction sequences is often governed by different considerations than those for lighter nuclei. This is especially important for experimental studies which not only may encounter additional problems when studying heavy nuclei but also have to be adapted in order to extract the data actually important for constraining nucleosynthesis. Nuclear theory also faces different challenges in heavy nuclei than in light ones and has to focus on the prediction of the actually relevant nuclear properties when applied to nucleosynthesis. Finally, a 3-D hydrodynamical simulation of exploding dense matter not only probes the limits of our nuclear models but also is a considerable computational challenge. The combined astrophysical and nuclear uncertainties lead to generally less well constrained conditions for nucleosynthesis in such phenomena and thus to considerable leeway in the interpretation and consolidation of astrophysical models.

\section{Differences to nucleosynthesis of lighter nuclei}
\label{sec:differences}

In this section, I want to highlight the special considerations that apply to the nuclear aspects of nucleosynthesis of intermediate and heavy nuclei. Some of the connected astrophysical uncertainties will be summarized in later sections.

\subsection{Stellar rates}
\label{sec:rates}

The connection between the astrophysical and the nuclear physics aspects of nucleosynthesis, in general, is made in the stellar reaction rate $r^*$, which for any reaction $a+A\rightarrow B+b$ is derived from the product of effective reaction area $F=n_A \sigma_{aA\rightarrow bB}^*(v,T)$ and the flux of incoming projectiles $J=n_b v(T)$, where $n_A$, $n_b$ are the number densities of target nuclei and projectile nuclei, respectively, in the stellar plasma with temperature $T$, and $v(T)$ is the relative velocity between the reaction partners $a$ and $B$. The stellar reaction cross section $\sigma_{aA\rightarrow bB}^*(v,T)$ also depends on $T$ because of the possible presence of nuclei in excited states -- and not only in the ground state (g.s.) -- due to thermal excitation in the plasma. The implications of this will be discussed in later sections. To obtain the number of reactions per volume and per second, an integration over all relative velocities has to be performed, where the probability to find particles at a velocity $v$ is given by the Maxwell-Boltzmann distribution (MBD) $\Phi_\mathrm{MB}(v,T)$ and thus
\begin{equation}
\label{eq:stellrategeneral}
r^*(T)=\int_0^\infty J(v,T)F(v,T) \,d^3v = n_a n_A \int_0^\infty \sigma_{aA\rightarrow bB}^*(v,T) v \Phi_\mathrm{MB}(v,T) \, d^3v \quad.
\end{equation}
Using center-of-mass (c.m.) coordinates, with $m_{aA}$ being the reduced mass, and inserting the MBD explicitly, this can also be rewritten as integration over the c.m.\  energy $E$ when realizing that $d^3v=4\pi v^2 dv$ and $E=mv^2/2$,
\begin{eqnarray}
\frac{r^*(T)}{n_a n_A} &= &\int_0^\infty \sigma_{aA\rightarrow bB}^*(v,T) v \left(\frac{m_{aA}}{2\pi kT}\right)^{3/2} e^{-m_{aA}v^2/(2kT)}\,d^3v \nonumber \\
&= &\sqrt{\frac{8}{\pi m_{aA}}} \left(kT\right)^{-3/2} \int_0^\infty \sigma_{aA\rightarrow bB}^*(E,T) E e^{-E/(kT)}\,dE \nonumber \\
&=& \int_0^\infty \sigma_{aA\rightarrow bB}^*(E,T) \Phi_\mathrm{plasma}(E,T) \,dE = \left< \sigma^* v \right>_{aA\rightarrow bB} \quad,
\label{eq:stellrate}
\end{eqnarray}
which allows to use the more common form of an energy-dependent reaction cross section $\sigma_{aA\rightarrow bB}^*(E,T)$ (see also Eqs.\ \ref{eq:effcs}, \ref{eq:stellcs}).\cite{fow74,raureview,hwfz} Instead of number densities $n$, astrophysical investigations often use abundances $Y=n/(\rho N_\mathrm{A})$, which implicitly contain the plasma matter density $\rho$ and Avogadro's number $N_\mathrm{A}$.

In explosive burning, however, it is not always necessary to determine $r^*$ at all temperatures. As introduced above, explosive burning involves high temperatures and sometimes also high densities. Typical temperatures and densities depend on the actual site and process but relevant temperatures, at which reaction rates have to be known, do not exceed 3.5 GK because forward and reverse rates are in equilibrium when the plasma is hotter. This means that the timescales for both forward and reverse reaction are short enough compared to the process timescale (the time during which temperature and density remain at values allowing the reactions to significantly modify the abundances) to permit the abundances of all involved nuclei to reach their equilibrium values, given by \cite{raureview}
\begin{equation}
\frac{Y_AY_a}{Y_FY_b}=\frac{g_0^A g_a}{g_0^F g_b} \frac{G^A_0}{G^F_0} \left( \frac{m_{Aa}}{m_{Bb}}\right) ^{3/2}e^{-Q_{Aa}/(kT)}
\end{equation}
for a reaction $A+a \leftrightarrow F+b$ and
\begin{equation}
\label{eq:equilcap}
\frac{Y_AY_a}{Y_C}=\frac{g_0^A g_a}{g_0^C} \frac{G^A_0}{G^C_0} \left( \frac{m_{Aa}kT}{2\pi \hbar^2}\right)^{3/2} e^{-Q_{Aa}/(kT)}
\end{equation}
for a reaction $A+a \leftrightarrow C+\gamma$ (see also Eq.\ \ref{eq:revphoto}), where $g_0^x=2J_0^x+1$ are the g.s.\ spin factors, $G_0^x=G^x/g_0^x$ the normalized partition functions for a nuclide $x$, and $Q_{Aa}$ the reaction $Q$ value when starting with $A+a$. Note that this does not imply that the abundances remain constant, they still depend on temperature $T$ which may vary with time as well as on $Y_a$ and $Y_b$, which also depend on the plasma density $\rho$.
The fact that individual rates do not appear anymore implies that the rates have only to be determined up to about 3.5 GK while the reaction $Q$ values (derived from nuclear masses) determine abundances at higher temperature. It should further be noted that not all reactions may be in equilibrium for given conditions. If all reactions were in equilibrium, this would lead to full \textit{nuclear statistical equilibrium} for which all abundances can be computed from coupled equations without the use of rates.\cite{raureview} On the other hand, any subset of reactions may be in equilibrium (e.g., neutron or proton captures and their reverse reactions) and the above equations can then be used to calculate the abundances affected by these reactions (which will be the fastest reactions occuring and thus be dominating).

The lower temperature limit is not defined as clearly as it depends much more on reaction cross sections, process duration, and the temperature and density profiles obtained in the astrophysical models. Generally, charged particle reactions freeze out at higher temperatures than reactions with neutrons due to the Coulomb barrier and are mostly too slow to change abundances considerably at $T<1$ GK.

Equilibrium considerations limit the temperature range in which astrophysical rates have to be known. Since the relevant energy window is determined by the energies giving non-negligible contributions to the integral in Eq.\ \ref{eq:stellrate} for a chosen temperature $T$, the thermal energies of the interacting nuclei are low by nuclear physics standards. These energies are below 10 MeV for charged particles and up to a few 100 keV for reactions with neutrons.\cite{energywindows} Especially for charged particle reactions on trans-iron nuclei, this implies tiny reaction cross sections which pose a significant challenges to their experimental determination.

\subsection{Impact of the nuclear level spacing}
\label{sec:nld}

Of great importance in the nucleosynthesis of trans-iron nuclides is the fact that heavier nuclei have a higher nuclear level density (LD) and thus smaller level spacing than light species.
This intrinsic LD is important in three ways. Firstly, the LD at the formation energy of the compound nucleus (which is the sum of incident projectile energy and separation energy of the projectile in the compound nucleus, $E_\mathrm{form}=E_\mathrm{c.m.}^\mathrm{proj}+S_\mathrm{proj}$) determines the dominant reaction mechanism. In the absence of levels to be populated close to the given energy, \textit{direct reactions} to low-lying final states dominate the reaction cross section.\cite{raureview,MaMe83,drc} When there is a small number of well separated excited states close to $E_\mathrm{form}$ in the compound nucleus $A+a$, this gives rise to resonances in the reaction cross sections, which can be described by the Breit-Wigner formula or by the R-matrix method.\cite{raureview,lt58,ilibook} The challenge therein is to determine the properties of the resonances contributing to the reaction rate integral in Eq.\ (\ref{eq:stellrate}).\cite{goerres} It is experimentally challenging to perform the required measurements for unstable nuclei and/or at low energy and theoretical \textit{ab initio} methods cannot be applied to heavy nuclei, yet. An extreme case of resonant reactions appears when the LD is high, leading to a large number of unresolved resonances which can be described by averaged resonance parameters. This is called the Hauser-Feshbach approach.\cite{haufesh,adndt} The vast majority of reactions in nucleosynthesis of intermediate and heavy nuclei can be described in this model.\cite{rtk} Despite of the increased number of nuclear transitions to be included in the model, this facilitates the predictions somewhat, as averaged quantities can be used. Calculations of reaction rates from smooth Hauser-Feshbach cross sections are also somewhat more ``forgiving'' to fluctuations around the ``true'' cross section value because of the integration over the projectile energy distribution (see Eq.\ \ref{eq:stellrate}). Different quantities, though, have to be known as input to the calculations than for direct or resonant reactions, such as optical potentials (related to the effective interaction in a many-nucleon system) and level densities (but not, e.g., spectroscopic factors of isolated levels).\cite{raureview}
Exceptions are reactions with very low or negative $S_\mathrm{proj}$ because then $E_\mathrm{form}$ is shifted to very low excitation energies at which the LD may be too low to apply the Hauser-Feshbach model even for heavier nuclei. Typically this occurs close to the driplines. Due to the lower LD at magic nucleon numbers, the Hauser-Feshbach model may not be applicable for magic target nuclei and low plasma temperatures. This applies mainly for neutron captures, however, because the projectile energy range relevant for the calculation of the reaction rate is shifted to higher compound excitation energies when there are charged particles in entrance or exit channel.\cite{energywindows}

The second consequence of a high LD is the availability of a larger number of states at low excitation energy which can be thermally populated in the stellar plasma. Temperatures in explosive burning are higher than in hydrostatic burning and forming nuclei at higher excitation energies. Combining the higher LD with higher $T$ has the consequence that the calculation of the \textit{stellar} rate includes a larger number of reactions on excited states of the target nucleus than would be the case for lighter nuclei, for which excited state contributions are often negligible. In fact, most reactions in trans-iron nucleosynthesis processes require to include more nuclear transitions in the calculation of reaction rates than just the ones originating from the g.s.\ of the target nucleus. The contribution of the rate $r_i$ obtained for a state $i$ with spin $J_i$ and excitation energy $E_i$ to the stellar rate $r^*$ can be quantified as \cite{stellarerrors,xfactor}
\begin{equation}
\label{eq:xfactor}
X_i(T)=\frac{w_ir_i(T)}{r^{*}(T)}=\frac{2J_i+1}{2J_0+1}e^{-E_i/(kT)}\frac{\int\sigma_i(E)\Phi_\mathrm{plasma}(E,T)dE}{\int\sigma^{\mathrm{eff}}(E)\Phi_\mathrm{plasma}(E,T)dE} \quad.
\end{equation}
The statistical weights $w$ are given by \cite{fow74}
\begin{equation}
\label{eq:statweight}
w_{i}(T)=\frac{\left(2J_{i}+1\right)\exp\left(-E_{i}/(kT)\right)}{\sum_{m}\left\{ \left(2J_{m}+1\right)\exp\left(-E_{m}/(kT)\right)\right\} }=\frac{\left(2J_{i}+1\right)\exp\left(-E_{i}/(kT)\right)}{G(T)}\quad.
\end{equation}
The reaction cross section $\sigma_i$ of state $i$ is defined as
\begin{equation}
\label{eq:statecs}
\sigma_i(E)=\sum_{j}\sigma^{i\rightarrow j}(E-E_i)
\end{equation}
and the effective cross section as
\begin{equation}
\label{eq:effcs}
\sigma^{\mathrm{eff}}(E)=\sum_{i}\sum_{j}W_i\sigma^{i\rightarrow j}(E-E_i)=G_0(T) \sigma^*(E,T) \quad,
\end{equation}
where $\sigma^{i\rightarrow j}$ is the partial cross section from
target level $i$ to final level $j$ (cross sections at zero or negative energies are zero).\cite{fow74,hwfz}
The modified weights $W_i$ are specified in Eq.\ (\ref{eq:new}).
For the g.s., Eq.\ (\ref{eq:xfactor}) simplifies to \cite{xfactor}
\begin{equation}
\label{eq:gsxfactor}
X_0(T)=\frac{\int\sigma_0(E)\Phi_\mathrm{plasma}(E,T)dE}{\int\sigma^\mathrm{eff}(E)\Phi_\mathrm{plasma}(E,T)dE} \quad.
\end{equation}
It is very important to note that this is different from the simple ratio $r_0/r^*$ of g.s.\ and stellar rates, respectively.\cite{stellarerrors} (The ratio $r^*/r_0$ is called the \textit{stellar enhancement factor} and in the past was mistakenly assumed to quantify the excited state contributions.)
Using the above equation, the total excited state contribution
\begin{equation}
\label{eq:exccontrib}
X_\mathrm{exc}=1-X_0=\sum_{i>0}^{n} X_i + \sum_{J\pi} \int_{E_n}^\infty \rho_\mathrm{LD} (E_\mathrm{exc},J,\pi) X(E_\mathrm{exc},J,\pi)\, dE_\mathrm{exc}
\end{equation}
can be defined, using a sum over discrete excited states and an integration over a LD $\rho_\mathrm{LD}$ above the last included excited state at energy $E_n$, with generalized $X(E_\mathrm{exc},J,\pi)$ depending on excitation energy $E_\mathrm{exc}$, spin $J$ and parity $\pi$, analogous to Eq.\ (\ref{eq:xfactor}).
In light nuclei, where the energy of the first excited state is larger than $kT$, $X_0 \approx 1$ and thus $X_\mathrm{exc} \approx 0$. Then $\sigma^*(E,T)\approx \sigma^*(E)=\sigma_0(E)$. On the other hand, $X_0\ll 1$ and $X_\mathrm{exc}\approx 1$ for heavy nuclei already at low plasma temperatures. A straightforward measurement of a reaction with the target nucleus being in the g.s.\ will therefore only constrain a small fraction of the stellar rate. Even when experimentally determining several transitions, one has to be aware that many more transitions contribute to rates involving heavier nuclei than for light nuclides. This is especially pronounced for charged particle reactions but it was shown that already neutron capture rates on nuclei in the rare-earth region exhibit sizeable excited state contributions even at $kT<30$ keV.\cite{stellarerrors}

Finally, a high LD alters the determination of resonance widths for resonant reactions and averaged widths in the Hauser-Feshbach model. In light nuclei, only few nuclear transitions $i\rightarrow j$ between a small number of states determine the reaction cross sections. Even when the cross sections cannot be measured directly, often indirect methods allow to determine the few dominating transitions and thus to derive the cross sections and rates. The situation is different for nuclei with high LD, as many more transitions contribute. Thus, resonance widths comprising all these transitions have to be measured or calculated. The utility of indirect methods or of studying reverse reactions is much smaller. For example, (d,p) and (d,n) reactions frequently used to study neutron and proton states, respectively, and their spectroscopic factors in light nuclei are of limited use here, as the number of levels at $E_\mathrm{form}$ is too large when the Hauser-Feshbach model is applicable. Also, the (d,p) and (d,n) cross sections themselves do not provide further information on the astrophysical rates unless they are applied to systems with low LD, for which the direct reaction mechanism dominates. They can be used, however, to identify low-lying isolated states which are of significance in the calculation of particle widths. Similarly, Coulomb dissociation is a helpful tool in studying light nuclei and can also be applied at radioactive ion beam facilities\cite{naka09}  but for heavier nuclei only a small subset of the relevant transitions can be accessed by this method. The utility of indirect methods to study rates involving heavier nuclei is further decreased by the larger excited state contributions $X_\mathrm{exc}$ to the stellar rate, implying even more contributing transitions, as discussed above. On the other hand, theoretical predictions are simpler for averaged quantities entering the prediction of widths, such as the compound nucleus LD in the calculation of $\gamma$ widths, and do not require \textit{ab initio} models but only effective theories.

\subsection{Sensitivities}
\label{sec:sensi}

Since systems with high LD exhibit different dependences of the cross sections on nuclear properties, an important quantity to consider is the sensitivity $s$ of cross sections or rates $\Omega$ to a change in an input quantity $q$, \cite{sensi}
\begin{equation}
\label{eq:sensi}
s=\frac{v_\Omega-1}{v_q-1} \quad.
\end{equation}
It is a measure of a change by a factor of $v_\Omega=\Omega_\mathrm{new}/\Omega_\mathrm{old}$ in $\Omega$ as the result of a change in the quantity $q$ by the factor $v_q=q_\mathrm{new}/q_\mathrm{old}$, with $s=0$ when no change occurs and $s=1$ when the final result changes by the same factor as used in the variation of $q$, i.e., $s=1$ implies $v_\Omega=v_q$.
This is equivalent to writing
\begin{equation}
s=\frac{q_\mathrm{old}}{\Omega_\mathrm{old}}\frac{d\Omega}{dq} \quad,
\end{equation}
with $d\Omega=\Omega_\mathrm{new}-\Omega_\mathrm{old}$ and $dq=q_\mathrm{new}-q_\mathrm{old}$, as used in standard sensitivity analysis.

Comparing the sensitivities of laboratory cross sections (for nuclei in their g.s.) and stellar reaction rates allows an improved judgement of the utility of a given experiment to constrain the rate or some relevant input.\cite{sensi} Since $s$ is strongly energy dependent, a measurement outside the energy range mainly contributing to the rate may not provide the desired, astrophysically relevant information. Different sensitivities of laboratory cross sections and rates also reflect the contribution of reactions on excited states to the stellar rate. Such reactions on excited states proceed at smaller relative energy and thus have a different sensitivity than those on the g.s.\ of a nucleus. In this context it should be noted that most determinations of relevant energies at which cross sections contribute significantly to the reaction rate, the \textit{Gamow window}, only provide them for the g.s.\ cross sections $\sigma_0$ but when $X_0\ll 1$ the relevant energies are much lower for the excited state contributions dominating the rate.

The sensitivity $s$ is also central to the assessment of the remaining uncertainties in a rate after the inclusion of experimental data. This will be further discussed in Section \ref{sec:sproc} below.
In order to determine the actual rate uncertainties it is important to not only know the experimental error (or uncertainty factor) but also to have an initial estimate of the theoretical uncertainty involved. Would we know the uncertainty in the prediction of, say, the averaged widths, the sensitivities defined above could be directly used to determine the resulting uncertainty in the stellar rate (or reaction cross section). Unfortunately, this poses a fundamental problem because theoretical uncertainties are fundamentally different from experimental errors, at least when talking about statistical errors in data. Several factors contribute to the theoretical uncertainty. First, an appropriate model has to be chosen, for example, by choosing which reaction mechanism to treat and in which way. Obviously, it is \textit{a priori} impossible to exactly determine the correctness of a model without being able to falsify it through comparison with data. This impossibility poses the fundamental problem of quantifying uncertainties of the theory approach.\cite{sensi} It is loosely related to systematic errors in experiments due to instruments and external effects, which are also hard to estimate. Assuming that a "correct" model is used, the next step in the uncertainty analysis would be to check the reliability of the input required to perform calculations with the model. Some of the input may be data bearing some measurement error. It is possible to propagate this error into the final model output, either analytically or, more likely, by variation studies, for example by a Monte Carlo variation of such input values within their errors. Part of the input, however, may again be in the form of a model (e.g., the LD or an energy-dependent $\gamma$-strength function). In this case the initial problem of determining the correctness of a model returns. Even when several choices for model descriptions of a quantity are available, the uncertainty cannot be determined by switching from one description to the other and observing the change in the result because this cannot be considered drawing a random sample from a statistically distributed set. The different treatments may even yield similar results at one energy but vastly differ at another. Without further knowledge of the underlying model assumptions it is even impossible to sensibly compare different approaches. Therefore it is fundamentally impossible to specify an error bar for a model, especially not a statistically distributed one. Theory uncertainties have to be estimated in comparison to available data but cannot be as strictly defined as statistical errors in experimental data. It is in this sense that the term ``theory uncertainty'' is used below. Monte Carlo variation of input can therefore only determine part of the full uncertainty, specifically the uncertainty stemming from input data (including natural constants).

In addition to uncertainty analysis, the sensitivity can also be used to directly infer the impact of a (experimentally or theoretically) newly determined quantity $q$ (for example, an averaged width in a Hauser-Feshbach calculation) on the cross section or reaction rate. With $\Omega_\mathrm{old}$ being the previous value of the cross section or rate of interest, the new value $\Omega_\mathrm{new}$ is simply given by
\begin{equation}
\Omega_\mathrm{new}=\Omega_\mathrm{old} \left(s\left(v_q-1\right)+1\right) \quad,
\end{equation}
when $v_q$ is the factor by which the newly determined $q$ differs from its previous value used to calculate $\Omega_\mathrm{old}$.

\section{Nucleosynthesis processes}
\label{sec:nucleosynthesis}

In the following I want to discuss relevant nuclear and astrophysical uncertainties in selected nucleosynthesis scenarios. The list of topics and challenges is not exhaustive but rather is supposed to provide examples which enable the reader to generalize and apply the considerations when dealing with further details and other nucleosynthesis processes.

\subsection{s-Process}
\label{sec:sproc}

The astrophysical slow neutron-capture process ($s$ process) produces about half of the elemental abundances between Fe and Bi.\cite{b2fh,al}
The $s$ process is attributed to environments of neutron number densities $n_n$ of typically $10^6-10^{12}$ cm$^{-3}$.\cite{kapgall} When an unstable nucleus is produced by neutron capture, $\beta$ decays are usually faster than subsequent neutron captures, so the reaction path follows the valley of stability. The $s$ process takes place
in different stellar sites. In particular, the $s$-process abundances in the solar system are made by contributions from different generations of stars, providing three different $s$-process components, a main, a weak and a strong component.\cite{kapgall}

The main component dominates the $s$ contributions to the region between Zr and Pb. It
is mainly associated with thermally pulsing Asymptotic Giant Branch (AGB) stars of $1-3$ $M_\odot$ with an initial
composition close to solar.\cite{gall98,arl99}
During the AGB phase, He burning takes place in a shell surrounding the C/O core of the star. Thermal pulses are caused by He-shell flashes which occur because He burning cannot sustain hydrostatic equilibrium within a thin shell. As a consequence of the mixing processes during the thermonuclear flashes, protons can be mixed downwards from the H-shell into the He-shell where they are used to produce $^{13}$C from $^{12}$C via $^{14}$N in the interpulse phase. The He-shell contracts and heats up before igniting another He-shell flash. This allows neutrons to be released through $^{13}$C($\alpha$,n) before and in such a shell flash.\cite{booth} During the temperature peak neutrons are further released in $^{22}$Ne($\alpha$,n) reactions. The strong component also originates in AGB stars but old ones, with metallicities much lower than solar.\cite{tra01} It is responsible for about half of the solar $^{208}$Pb abundance and for the $s$-process abundance of Bi. The astrophysical challenge in the investigation of the $s$ process in AGB stars is in the detailed modelling of the downmixing of protons from the H-shell into the He-shell and the formation of a $^{13}$C ``pocket''. The convection and mixing can only be understood as a 3D effect but a fully 3D stellar model including nucleosynthesis is computationally not yet feasible. Therefore all AGB $s$-process nucleosynthesis models include \textit{ad hoc} assumptions and phenomenological parameters describing the $^{13}$C pocket and it is not clear how these evolve for stars with different masses and compositions.

The weak $s$ process takes place in massive stars (with more than 8 $M_\odot$) which later explode as a core-collapse supernova, and is producing most of the $s$ abundances in the mass region between Fe and Zr.\cite{kapgall,rhhw02} In these stars, neutrons are mostly produced at the end of core He-burning and during the later convective C-shell burning phase via $^{22}$Ne($\alpha$,n). Typical neutron energies are $kT=0.008$ and 0.03 MeV for interpulse and flash burning and around 0.09 MeV in massive stars. (Note that the relevant plasma temperature in GK ($T_9$) is then given by $T_9=11.6045kT$ when $kT$ is in MeV.)

On the nuclear physics side, the challenges lie in the accurate determination of the neutron producing and consuming rates.
In the last decades, considerable effort has been put into the measurement of precise and accurate neutron capture cross sections at $s$-process energies.\cite{bao,kadonis} It has been understood that the uncertainties in the present neutron capture rates are dominated by the experimental errors which are currently below 5\% for most $s$-process captures.\cite{wiekapplan} Recently it has been shown, however, that even at the comparatively low $s$-process temperatures excited state contributions $X_\mathrm{exc}$ are not negligible, especially for deformed nuclei.\cite{stellarerrors} This leads to larger remaining uncertainties because the laboratory cross section only constrains a fraction of the stellar rate. The situation is even worse when the experimental ``rate'' is inferred not from neutron capture but its reverse reaction, ($\gamma$,n). Ground-state contributions $X_0$ are $3-4$ orders of magnitude smaller in ($\gamma$,n) rates than in (n,$\gamma$) ones at the same temperature. This will be further discussed in Section \ref{sec:gamma}.

Here, we take a look at an important example in the $s$ process, $^{185}$W(n,$\gamma$). The nucleus $^{185}$W is unstable and the location of a branching in the $s$-process path where neutron capture and $\beta$ decay are competing. The current version of the neutron capture compilation for the $s$ process (KADoNiS, v0.3) recommends an experimental value with a 9\% error at $kT=30$ keV.\cite{kadonis} Even when assuming that the uncertainties in the predictions of the transitions originating on the g.s.\ and the excited states are uncorrelated, this would suggest a strong experimental constraint on the rate because for $^{185}$W(n,$\gamma$), $X_0=0.98$ at $kT=8$ keV and $X_0=0.75$ at $kT=30$ keV. The new uncertainty factor $u^*$ of the stellar rate is constructed from the original theory uncertainty factor $U_\mathrm{th}$ and the experimental uncertainty factor $U_\mathrm{exp}$ by using \cite{stellarerrors}
\begin{equation}
u^*=U_{\mathrm{exp}}+(U_\mathrm{th}-U_{\mathrm{exp}})X_\mathrm{exc}=U_{\mathrm{exp}}+(U_\mathrm{th}-U_{\mathrm{exp}})(1-X_0) \label{eq:uncertainty} \quad.
\end{equation}
The 9\% experimental error would translate into $U_\mathrm{exp}=1.09$. Due to the large $X_0$, even a sizeable theory uncertainty of a factor of 2 or more would be almost completely removed. A closer inspection of the source of the quoted rate, however, reveals that the recommended value is obtained by averaging the rates given by two photodisintegration experiments\cite{sonn03,mohr04}, measuring $^{186}$W($\gamma$,n)$^{185}$W cross sections. To arrive at a rate, the ($\gamma$,n) cross section of the g.s.\ of $^{186}$W was compared to a Hauser-Feshbach prediction.\cite{sonn03,mohr04} It was found that the prediction had to be scaled to reproduce the data. Consequently, the (n,$\gamma$) rate predicted by the same model was scaled by the same factor and this value was used as ``experimental'' rate. The quoted error actually applies to the ($\gamma$,n) cross section. Therefore it has to be interpreted differently when asking for the remaining uncertainty in the neutron capture rate.

\begin{figure}[t]
\begin{center}
\includegraphics[width=1.1\textwidth]{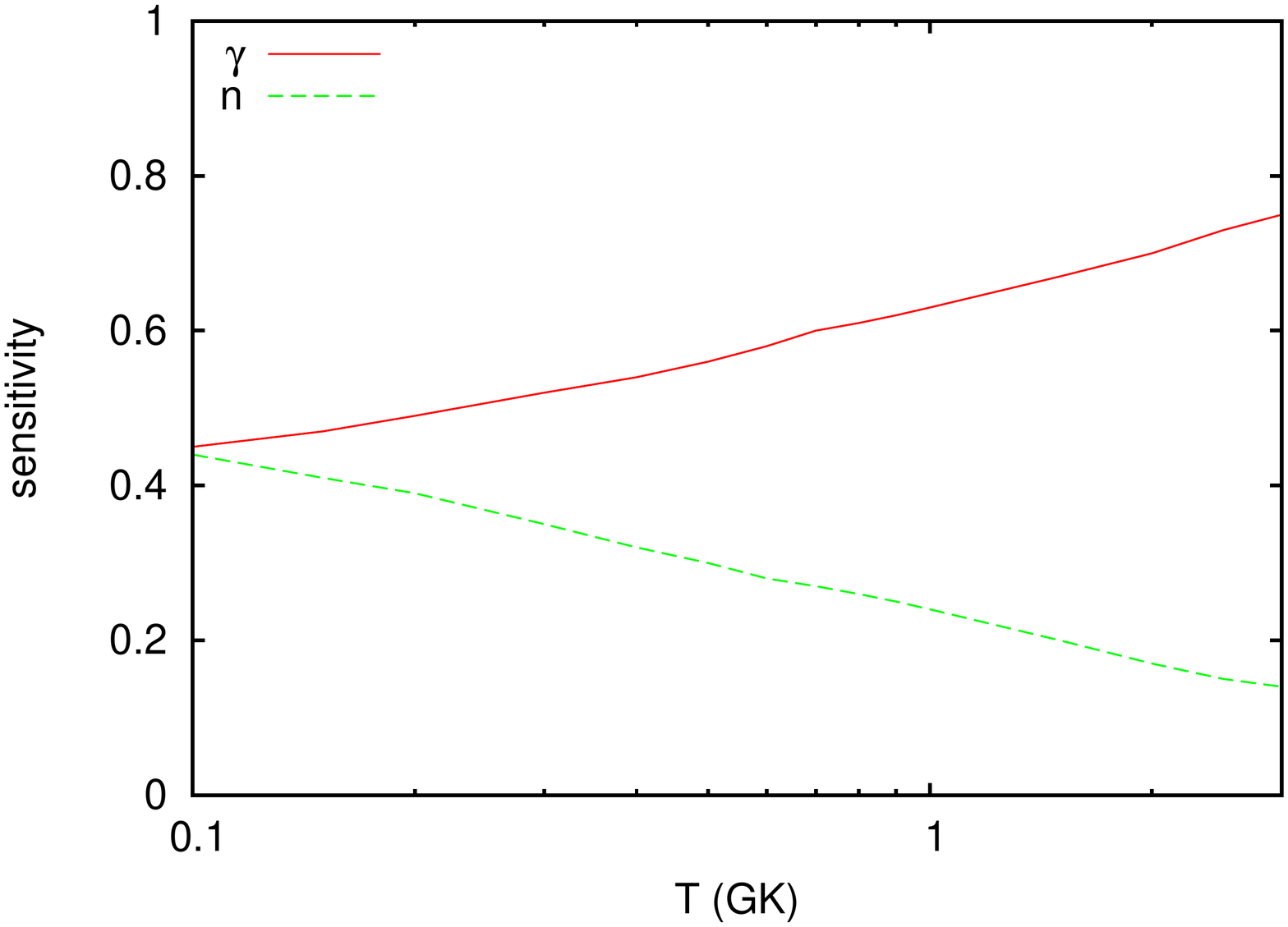}
\caption{Absolute values of the sensitivity $|s|$ of the stellar $^{185}$W(n,$\gamma$) rate to variations in the neutron and $\gamma$ widths as function of plasma temperature.\label{fig:w185}}
\end{center}
\end{figure}

If we could constrain the reverse \textit{stellar} rate well, the forward rate could be computed equally reliably because of the reciprocity relation for \textit{stellar} rates. Based on the detailed balance theorem for nuclear reactions, it relates the rates of the forward and reverse reactions and reads
\begin{equation}
r^*_{C\gamma}=\frac{g_0^A g_a}{g_0^C} \frac{G^A_0}{G^C_0}
\left( \frac{m_{Aa}kT}{2\pi \hbar^2}\right)^{3/2} e^{-Q_{Aa}/(kT)} \left< \sigma^* v \right>_{aA\rightarrow bB} \label{eq:revphoto}
\end{equation}
for a photodisintegration reaction $C+\gamma \rightarrow A +a$ (see also Eqs.\ \ref{eq:stellrate}, \ref{eq:equilcap}).\cite{raureview}

Equation (\ref{eq:revphoto}) only applies to \textit{stellar} rates in both the forward and reverse channel, i.e., it can only be used with a rate or cross section of a nucleus in the g.s.\ -- as usually encountered in laboratory measurements -- when $X_\mathrm{exc} \ll 1$ (as defined in Eq.\ \ref{eq:exccontrib}), and thus $\sigma_0 \approx \sigma^\mathrm{eff}$ and $G_0\approx 1$. For $^{186}$W($\gamma$,n), however, $X_0=7\times 10^{-3}$ at $T_9=0.1$ and $X_0=5\times 10^{-3}$ at $T_9=0.4$, for instance.\cite{sensi} It immediately follows that the measured contribution to the stellar rate is actually negligible. Does the measurement at least provide information on a transition that is important for the neutron capture cross section? The $X_0$ for the photodisintegration are closely related to the relevance of the measured $\gamma$ transition to/from the g.s.\ of $^{186}$W. Among the $\gamma$s emitted in the capture, the ones directly to the g.s.\ are only a tiny fraction of all. The majority are emitted with energies of $2-4$ MeV,\cite{gammaenergies} about half-way between the g.s.\ and the neutron separation energy (see also Section \ref{sec:gamma}). Thus, the ($\gamma$,n) experiment probes the $\gamma$-strength function at an energy which is important neither for the capture nor for the stellar ($\gamma$,n) rate. And finally, even if the $\gamma$ width was constrained, this would still not fully constrain the (n,$\gamma$) rate. As shown in Fig.\ \ref{fig:w185}, the rate is equally sensitive to uncertainties in the $\gamma$- as well as the neutron width at low temperatures.

From the above considerations it follows that the $^{185}$W(n,$\gamma$) rate is experimentally still unconstrained at $s$-process temperatures and its measurement remains a challenge due to the half-life of $^{185}$W. (The limitations of photodisintegration experiments are more generally discussed in Sec.\ \ref{sec:gamma}.) From $s$-process models it is found that the solar $^{186}$Os/$^{150}$Sm abundance ratio, which is determined by the branching at $^{185}$W, is best reproduced by a value which is 1.81 times larger than the averaged value of the two photodisintegration experiments.\cite{bisterzo} This is larger by a factor of 1.5 than the previously recommended value which is based on a Hauser-Feshbach prediction and is a reasonable theory uncertainty for such a neutron capture rate.\cite{rtk}

The above does not only apply to neutron captures in the $s$ process but similar investigations of contributions and sensitivities have to be performed for any reaction in any nucleosynthesis process, preferrably before the measurement is undertaken.

\subsection{r-Process}
\label{sec:rproc}

In order to explain the solar abundance pattern beyond Fe, another neutron capture process had to be postulated, the $r$ process.\cite{b2fh,al} The abundance peaks at magic neutron numbers at stability are accounted for in the $s$ process, due to diminished neutron capture cross sections of neutron-magic nuclei. The existence of additional peaks shifted to lower mass number was explained by neutron captures far off stability, where neutron-magic nuclei are encountered at lower mass number.\cite{bsmey94,wallerstein,ctt91,arngorr} This requires neutron number densities larger than 10$^{20}$ cm$^{-3}$ and temperatures around 1 GK.\cite{ctt91,frei} At such conditions, (n,$\gamma$)-($\gamma$,n) equilibrium is established within each isotopic chain. Applying Eq.\ (\ref{eq:equilcap}) to neutron captures, it can be shown that the abundances within such a chain vary quickly with neutron separation energy $S_\mathrm{n}$ and only one or two isotopes receive non-negligible abundances. Because $\beta$-decays are slower than the equilibrated reactions, even far from stability, the isotope with the maximal abundance in each chain is called \textit{waiting point}. The next isotopic chain is reached only by decay of the waiting point nucleus.\cite{frei,doesitmakedifference} Since the waiting points in each chain are disconnected from each other and equilibrium is reached almost instantaneously (on a timescale on the order of reaction timescales of 10$^{-16}$ s or less), there is no well-defined reaction ``path'' as in the $s$ process. An $r$-process path is sometimes shown, nevertheless, as an imaginary line connecting the waiting points of all isotopic chains. But also the fact that the $r$-process peaks are wider than the $s$-process ones shows that even this type of path, i.e., the waiting points, is not as well defined as the $s$-process path.

The waiting points are located far from stability and therefore their half-lives are short, a few 100 ms. The total processing time to go from $Z=26$ to $Z=83$ and beyond is on the order of seconds. This short timescale along with the high neutron densities and high temperature indicates an explosive environment. So far, the actual site of the $r$ process is unknown. The long time favorite have been the innermost, still ejected layers of a massive star in its core-collapse supernova (ccSN) explosion.\cite{bsmey94,ctt91,arngorr} These layers cool down from very high temperatures while moving in a neutrino wind emitted from the nascent neutron star. The neutrino interactions render the matter neutron rich, so that an $r$ process can ensue once the matter has cooled sufficiently from nuclear statistical equilibrium to allow neutron captures.\cite{thiele11,faru10} Recent multi-D models of core-collapse explosions, however, have been unable to provide the entropies required to explain the full solar $r$-process pattern.\cite{thiele11,arcoreview,thom01,wan06,arco07,fischer,nishimura} Additionally, it would be problematic to avoid overproduction at mass numbers below 80 when producing all three $r$-process peaks in a single core-collapse event.\cite{frei}

A promising alternative site are neutron star mergers which may allow nucleosynthesis in (n,$\gamma$)-($\gamma$,n) equilibrium in even neutron-richer matter than in core-collapse ejecta and thus require a smaller entropy range to produce solar abundances.\cite{merger} Since such mergers are rare events which occur late in the Galaxy, they cannot explain the $r$-process patterns found in very old stars, so-called ultra-metal poor stars, which formed early in the Galaxy.\cite{thiele11} These patterns, however, suggest that there may be two $r$-process components, a weak component contributing to nuclides below Ba and a stronger, robust component contributing to abundances of Ba and above.\cite{tra04,weak-r,cownature,honda} Independently, two components have been suggested based on meteoritic data.\cite{qian}

Further alternatives for possible sites are neutron-rich material in magnetically driven jets during ccSN, which circumvent the problems in reproducing solar abundances in neutrino-driven outflows.\cite{fuji,kaeppeli} A ``cold'' $r$-process neutrino-wind scenario was also postulated, in which (n,$\gamma$)-($\gamma$,n) equilibrium is not achieved.\cite{wan07,faru10}

The astrophysical uncertainties, i.e., the knowledge of mechanisms and conditions, are considerable, especially in neutrino-wind models. These are strongly coupled to the lack of our knowledge regarding the ccSN explosion mechanism. Further insights can only be brought by 3D hydrodynamics calculations with realistic neutrino transport to constrain the actual conditions, such as neutrino-wind properties (determining the evolution of the neutron-richness of the ejected material) as well as the behavior of the ejecta regarding fallback and wind-termination shock, which can alter the time evolution of the nucleosynthesis conditions considerably.\cite{thiele11,arcoreview}

The stellar reaction rates entering the (post-processing) nucleosynthesis calculations also bear considerable uncertainties. Due to the involvement of very neutron-rich nuclides far off stability, theory is probed at the edge of our knowledge. Experiments are still unfeasible, except for studies of masses and half-lives somewhat closer to stability, but future radioactive ion beam facilities may allow to gather further information on these nuclei and their reactions. They will have to overcome severe limitations on what kind of nuclear properties can be studied with respect to astrophysical applications. Due to the uncertainties in astrophysical sites and conditions, the range of nuclei to be studied is also large, much larger than, e.g., in the $s$ process with its well-determined path.

Contrary to the $s$ process, neutron capture rates, however, do not have to be known as long as (n,$\gamma$)-($\gamma$,n) equilibrium is valid.\cite{doesitmakedifference} The equilibrium abundances are solely determined by $S_\mathrm{n}$ which, in turn, is derived from nuclear mass differences. Individual neutron captures become important in the freeze-out from equilibrium (or in a cold $r$ process), when they compete with $\beta$-decays at $T<1$ GK. Depending on the freeze-out timescale, this may or may not impact the final abundances.\cite{doesitmakedifference} While the lower $r$-process peaks remain almost unaffected by the freeze-out (because it is fast for its progenitor nuclei), the third peak and the rare-earth abundances in between the second and third peak have been found to show some sensitivity to freeze-out.\cite{doesitmakedifference,surman,mumpower} The relevance of these sensitivities, however, is not yet clear because considerable contributions to the rare-earth region are expected from fission fragments of nuclei produced above Bi and these alone may robustly explain the rare-earth peak.\cite{gorfiss}
Individual neutron capture rates also determine how long (n,$\gamma$)-($\gamma$,n) equilibrium can be upheld while density and temperature are dropping.

Predictions and experiments are likewise complicated by the fact that during freeze-out the abundances move from nuclei with low $S_\mathrm{n}$ to such with higher $S_\mathrm{n}$ and therefore several reaction mechanisms may contribute, depending on the LD (see Sec.\ \ref{sec:nld}). Close to the neutron dripline and at magic neutron numbers far off stability, direct capture may be non-negligible although single resonances may still dominate the cross sections, depending on the level structure.\cite{raureview,MaMe83,drc,gordrc} Since the nuclei are heavy and their level structure is unknown, it is conceivable that isolated resonances already play a role even for the neutron-richest species appearing during $r$-process nucleosynthesis. Resonant capture definitely becomes an issue in later freeze-out phases and in a cold $r$ process. Resonant rates are extremely sensitive to the resonance energy, as it only contributes within the relevant energy range of the rate integral (Eq.\ \ref{eq:stellrate}), but it cannot be predicted with the necessary accuracy. Closer to stability, the Hauser-Feshbach model may already yield a good approximation to the rates, even at comparatively low LD, as long as the averaged resonance parameters used in this model are close to the average of the resonances within the astrophysically relevant neutron energy window (see also Sec.\ \ref{sec:nld}). Some attempts have been undertaken to combine predictions of direct capture rates with Hauser-Feshbach rates but the uncertainties are still considerable.\cite{raureview,gordrc} For experiments, resonant reactions pose a challenge because indirect methods (such as Coulomb dissociation but also photodisintegration, see Sec.\ \ref{sec:gamma}) rely on the fact that only few transitions are important for the calculation of the rate and this is only the case for direct reactions. Another problem are the elevated temperatures encountered in the $r$ process which lead to considerable population of excited states with all the consequences of additional transitions discussed before in Secs.\ \ref{sec:nld}, \ref{sec:sproc}.

Further required information includes rates for $\beta$-decay, $\beta$-delayed neutron emission, $\beta$-delayed and neutron-induced fission, and the fission fragment distributions. The temperature dependence for these rates (similarly to what was discussed before for neutron rates) has also to be taken into account. This has not even been consistently treated in theoretical rate predictions, yet.

\subsection{Origin of p-nuclides}
\label{sec:p-nucl}

Nucleosynthesis of elements heavier than iron is dominated by neutron capture processes. However, there are 35 proton-rich isotopes between Sr and Hg, which cannot be synthesized by the neutron-induced capture reactions of the \textit{s} and \textit{r} processes.\cite{p-review,arngorp} Their abundances are typically factors of $0.1-0.001$ smaller than $s$ and $r$ abundances. The bulk (32 nuclides) of these so-called \textit{p} nuclei are believed to be produced in a series of photodisintegration reactions, referred to as the $\gamma$ process, \cite{woohow} which is found in the outer layers of massive stars during their explosion in a core-collapse supernova \cite{woohow,rayet95,rhhw02} or in explosions of White Dwarfs as type Ia supernovae (SNIa).\cite{howmey,iwamoto,travaglio} At temperatures between 2 and 3 GK, a heavy seed distribution is first altered by ($\gamma$,n) reactions and thus proton-rich nuclei are created. After several neutron emissions, ($\gamma$,p) and ($\gamma$,$\alpha$) reactions start to compete with neutron emission, deflecting the reaction flux. The whole process lasts only a few seconds, before the environment becomes too cool to permit photodisintegration and $\beta$-decay chains finally populate the next stable isobar encountered for a given mass number. The seed nuclei either are $s$- and $r$-nuclei initially present (inherited by the massive star from its proto-stellar cloud or accreted from a binary companion in the case of White Dwarfs) or made in the object during its evolution ($s$ processing in the accreted layer on the White Dwarf or during the accretion).

A longstanding problem is posed by the comparatively large abundances, relative to the other $p$ nuclei, of $^{92,94}$Mo and $^{96,98}$Ru. They are not produced in massive stars as there is not enough $s$- and $r$-process material present to act as seed to produce these nuclides in sufficient quantities. Recently, it has been found that they can be produced in thermonuclear explosions of White Dwarfs which accreted highly $s$-process enriched matter from a companion AGB star.\cite{travaglio} Although this $s$-process enrichment is crucial for the success of the model, the details are uncertain and current $\gamma$-process studies in SNIa make \textit{ad hoc} assumptions because self-consistent simulations of $s$ processing in the accreted layer are difficult and unavailable to date. Another open question is in the details of the SNIa explosion, requiring multi-D hydrodynamical models which also provide the conditions for the nucleosynthesis in the explosion and the White Dwarf fragments. Finally, it remains unclear what fraction of SNIa is actually caused by mass accretion from a companion AGB star (single degenerate scenario) or by collisions of two White Dwarfs (double degenerate scenario).

A recent review discusses in detail the astrophysical and nuclear uncertainties when studying the origin of $p$ nuclei.\cite{p-review}
On the nuclear side, detailed reaction network calculations,\cite{rhhw02,rapp,raudeflect} have to consider several hundred reactions for which the reaction rates under the given conditions have to be known. Measurements are challenging or even impossible for several reasons: The cross sections at astrophysically relevant energies are very small, the g.s.\ contributions $X_0$ are tiny, and many potentially important reactions involve unstable target nuclei.\cite{p-review} Therefore experiments are limited to testing model predictions of certain transitions which are also appearing in the calculation of the astrophysical rate. Nevertheless, the experimental database of relevant reactions and nuclear properties for the $\gamma$ process is far from complete.\cite{p-review,kadonis} The majority of reaction rates has to be predicted by theory making use of the Hauser-Feshbach approach (see Sec.\ \ref{sec:nld}).

A crucial input parameter of the Hauser-Feshbach model are optical model potentials, describing the interaction between nuclei and light nuclear particles and required to determine particle transitions. Particularly large discrepancies have been found between predicted and measured $\alpha$-induced cross section data in the energy region of relevance.\cite{p-review,somo98,gledenov,gyur06}. This also has consequences for understanding isotope ratios in meteorites containing matter preserved from the early solar system.\cite{woohow90,rausm144,somo98,raucoulex} Therefore many experimental studies relevant to the production of $p$ nuclei have focused on better constraining the $\alpha$+nucleus potential. It is interesting to note that so far it was not possible to find a global optical potential which can consistently describe all the known ($\alpha$,$\gamma$) and ($\alpha$,n) data at low energy, especially when including the $^{144}$Sm($\alpha$,$\gamma$)$^{148}$Gd measurement below 12 MeV.\cite{somo98} This may require a complicated energy dependence of the optical potential parameters, fine-tuned to each nucleus separately.

Recently, a different solution to this $\alpha$-potential puzzle was proposed.\cite{raucoulex} It was suggested that an additional reaction channel may contribute to the total reaction cross section, which is not yet considered in the present Hauser-Feshbach models. Instead of modifying the energy-dependence of the optical potential, leading to a different total cross section, the new idea is that this total cross section is actually described well by the standard potential but that rather it has to be distributed differently between the possible reaction channels at low energy. Such an additional channel acting at subCoulomb energy (but being negligible at intermediate energies) is Coulomb excitation of target states. Test calculations have shown that this can consistently explain the $^{144}$Sm($\alpha$,$\gamma$) measurement and other low-energy data for nuclei with $N\geq 82$.\cite{raucoulex} Remaining uncertainties include the $B(E2)$ values (especially for odd nuclei) required for the calculation of the Coulomb excitation cross section as well as the knowledge of excited states contributing, especially for unstable nuclei.

\subsubsection{Can measurements of nuclear photodisintegration help to constrain astrophysical photodisintegration rates?}
\label{sec:gamma}

The fact that photodisintegrations play a role in certain astrophysical processes, specifically in the $\gamma$ process, often gives rise to the misconception that a photodisintegration measurement will directly help with determining the astrophysical reaction rate. Some of the problems when using photodisintegration reactions to determine stellar rates have already been addressed in Section \ref{sec:sproc}. Here I intend to provide further details. Full reviews of all relevant effects \cite{raureview} and photodisintegration in the $\gamma$-process \cite{p-review} are given elsewhere. Tables of sensitivities of cross sections and stellar rates to various nuclear properties and all g.s.\ contributions $X_0$ to the stellar rate are also available.\cite{sensi}

The main reason for the limited usefulness of photodisintegration experiments is that, even when exciting the nucleus to an astrophysically relevant energy (and this is important because at higher energies so many additional effects are coming into play that it becomes difficult to impossible to extract the ones relevant in the astrophysical energy range), the photon-induced transition from the g.s.\ with subsequent particle emissions is only a tiny fraction of all transitions appearing in the astrophysical case (see also Secs.\ \ref{sec:nld}, \ref{sec:sproc}).\cite{raureview,gammaenergies,raukiss,xfactor,sensi,stellarerrors,p-review,coulsupp}

The majority of the relevant $\gamma$-transitions have lower $\gamma$-energy (about $2-4$ MeV) and start at an already excited state of the nucleus (and therefore will probably also select different spin and parity quantum numbers).\cite{raureview,gammaenergies} The g.s.\ contribution $X_0$ to the stellar photodisintegration rate is always very small, often less than a percent.\cite{raureview,raukiss,mohr03,sensi,stellarerrors,p-review,rausanantonio} (A typical example is given in Sec.\ \ref{sec:sproc} for $^{186}$W($\gamma$,n) but the situation is worse for the $\gamma$-process because the higher plasma temperatures lead to even smaller $X_0$.) Therefore, it is always recommended to measure the capture because its g.s.\ contribution is larger and thus the rate can be better constrained by experiment (although some theory may still be necessary).\cite{raureview,raukiss,sensi,coulsupp}

(The situation may be different for nuclei with very low particle separation energy (or reaction $Q$ value) but these nuclei are far from stability and not yet accessible by experiments.\cite{gammaenergies,sensi})

Three components act together in making photodisintegrations less favorable: \cite{raureview}
\begin{enumerate}
\item The principle of detailed balance in nuclear reactions states that there is a simple relation between forward and reverse reaction between each discrete initial state $i$ and final state $j$. The cross sections $\sigma^{i\rightarrow j}$ are related by a simple phase space factor.
\item In astrophysics, most nuclei are in thermal equilibrium with the surrounding plasma and therefore are present also as excited nuclei, not just nuclei in their g.s. The fraction of nuclei in a specific excited state is given by a Boltzmann population factor depending on the energy of the excited state $E_i$ and the plasma temperature $T$: $P_i=(2J_i+1)\exp(-E_i/(kT))$. This leads to the probabilities $w_i$ introduced in equation (\ref{eq:statweight}).
\item The projectile (or $\gamma$) energies are also thermally distributed in a plasma. Obviously, the same energy distribution must act on a nucleus in the g.s.\ and one in an excited state. This implies that the respective energy distributions (for example, MBD or Planck distribution) are shifted relative to the integration energy $E$ relative to the g.s.\ and extending to higher excitation energy $E+E_i$ when acting on an excited state.\cite{fow74}
\end{enumerate}
It is important to realize that the above applies to all initial and final nuclei simultaneously in a reaction in an astrophysical environment. This is also why it is possible to derive the reciprocity relation between forward and reverse \textit{stellar} rates, as shown in Eq.\ (\ref{eq:revphoto}).
Already from the notion that a capture reaction (with positive $Q$ value) releases $\gamma$s to a wider range of final states than particles (e.g., neutrons) are released in its reverse (photodisintegration) reaction it can be seen that a larger number of $\gamma$ transitions are relevant than particle transitions (see also Sec.\ \ref{sec:sproc}).\cite{gammaenergies} From the reciprocity relation in Eq.\ (\ref{eq:revphoto}) it is easy to show that the number of participating transitions is larger in the exit channel of a reaction with positive $Q$ value also in stellar rates.\cite{raureview,sensi,stellarerrors}

Often it is confusing to only consider the Boltzmann population factors $P_i$ of the excited states as given above. They seem to give similar weights to low-lying excited states, regardless of which reaction direction they are applied to. It is incorrect, however, to take only these factors as a measure of the importance of initial states. Rather, folding these factors with the shifted energy distributions of the projectiles (or $\gamma$s) and expressing everything in an energy scale relative to the ground state leads to a transformed weight \cite{raureview}
\begin{equation}
\label{eq:new}
W_i(E)=(2J_i+1)(1-E_i/E) \quad.
\end{equation}
This can be seen in the full expression for the stellar cross section, \cite{raureview,rausanantonio}
\begin{eqnarray}
\sigma^*(E,T)&=&\frac{1}{G(T)} \sum_i \sum_j (2J_i+1) \frac{E-E_i}{E}
\sigma^{i \rightarrow j}(E-E_i) \nonumber \\
&=&\frac{1}{G(T)} \sum_i \sum_j W_i \sigma^{i \rightarrow j}(E-E_i)
\quad,
\label{eq:stellcs}
\end{eqnarray}
to be used in Eq.\ (\ref{eq:stellrate}) to calculate the stellar reaction rate.
This means that the partial cross sections for reactions on excited states are evaluated at $E-E_i$ instead of the usual $E$ and their contributions weighted by $W_i$. Cross sections for individual transitions $\sigma^{i \rightarrow j}$ are zero for negative energies.\cite{fow74} Equation (\ref{eq:stellcs}) applies to the forward as well as the reverse reaction.

This has two implications. Firstly, the $W_i$ are linear, more slowly declining than the exponentially declining Boltzmann factor. Secondly, the weight is relative to the integration energy $E$ in Eq.\ (\ref{eq:stellrate}) and thus energies in the Gamow window will contribute most. The Gamow window is at higher excitation energy of the target nucleus in a reaction with negative $Q$ value.\cite{energywindows} (When its energy is $E_\mathrm{G}$ in the capture, then it is $E_\mathrm{G}+Q_\mathrm{capture}$ in the photodisintegration.) Therefore, the weights decline more slowly and reach higher excitation energy in the case of photodisintegration than in the case of capture (see also Fig.\ 9 of Ref.\ \cite{raureview}).

Finally, folding the actual astrophysical weights $W$ with the relevant $\gamma$ transition strengths (which are - close to stability - roughly about halfway between the ground state and the particle separation energy \cite{gammaenergies}) shows that the ground state contribution to astrophysical photodisintegration rates is small.\cite{rausanantonio}

From the above, it is found that photodisintegration (with real photons or by Coulomb fragmentation) cannot directly constrain the astrophysical reaction rate for systems with large inherent LD and at high temperature, as found in explosive nucleosynthesis above Fe. Capture reactions are much better suited for this as their g.s.\ contribution to the stellar rate is much larger than for photodisintegrations. Scrutinizing the complete tables\cite{sensi} of g.s.\ contributions across the nuclear chart, it can be seen that \textit{stellar} photodisintegration rates almost always have extremely tiny g.s.\ contributions (with very few exceptions \cite{coulsupp}). Therefore a measurement provides a value closer to the rate required in astrophysics. Nevertheless, further corrections from theoretical calculations may still be needed.

Can photodisintegration experiments help at all? From the above, it can be seen that a ($\gamma$,n) experiment tests the $\gamma$-ray strength function (or $\gamma$ width) at much larger $\gamma$-ray energy than the one of the $\gamma$-rays actually contributing mostly to the \textit{stellar} photodisintegration rate.\cite{gammaenergies} In the past, a few ($\gamma$,n) cross sections were measured\cite{sonn03,mohr00,mohr03,mohr04} and an astrophysical rate was derived by renormalizing the predicted rate by the same factor as found when comparing the measured result with a prediction in the same model as used to calculate the rate (as discussed in Sec.\ \ref{sec:sproc}). This implicitly assumes that any discrepancy found between experiment and theory for the much larger $\gamma$ energy of the g.s.\ photodisintegration equally applies also to the actually relevant transitions with smaller relative energy (or are due to the particle transitions in the exit channel). Usually, this is not a good assumption because the photon strength function behaves differently at low energy and the allowed partial waves in the particle transitions may have different relative angular momentum than the ones appearing in the astrophysical rate. Therefore the experiment does not constrain the astrophysical rate.

Studying the behavior of the $\gamma$-strength function itself aids the prediction of astrophysical rates, both for captures and photodisintegrations. As mentioned above, the relevant $\gamma$ energies are of the order of $2-4$ MeV.\cite{gammaenergies} Changes in the strength function within this energy range directly affect astrophysical capture and photodisintegration rates, changes outside this range are of smaller importance.\cite{gammaenergies,p-review} Unfortunately, this $\gamma$-energy range can only be accessed indirectly, requiring a combination of experiment and theory. For example, ($\gamma$,$\gamma'$) data can help to constrain the low-energy $\gamma$ strength or -width. The ($\gamma$,n) data cannot be used for this by themselves. Recent experiments\cite{utso13} realized this and fit experimental ($\gamma$,$\gamma'$) data with theoretical models to use such models for the calculation of the stellar photodisintegration rate. Note that in such experiments also any ($\gamma$,n) cross section, if measured, does not contribute significantly to the actual constraint of the stellar rate.

Because of the relevance of reactions proceeding on thermally excited target states in an astrophysical plasma, it is important to study (particle) transitions from these excited states. This can be done by studying the reverse transitions to such excited states in a particle exit channel. Photodisintegrations can achieve this, as can inelastic particle scattering (e.g., (n,n')).\cite{mosc10,stellarerrors,p-review} One has to be careful in the interpretation of such experiments, however, because starting from a specific ground state introduces a selection of possible quantum numbers (relative angular momenta) which may be different from the astrophysically relevant ones. Nevertheless, this may allow testing the predictions of particle transitions to/from excited states by theoretical models and also the predicted ratios to the ground state transitions (e.g., ($\gamma$,n$_1$)/($\gamma$,n$_0$), ($\gamma$,n$_2$)/($\gamma$,n$_0$), \dots; ($\gamma$,$\alpha_1$)/($\gamma$,$\alpha_0$), ($\gamma$,$\alpha_2$)/($\gamma$,$\alpha_0$), \dots). \cite{p-review}

(Additional information required in predictions of captures and photodisintegrations are low-lying discrete states and nuclear level densities above those discrete states.)

What is essential in all types of experiments is to really measure within the relevant energy region. The $\gamma$ strength function has to be known around $2-4$ MeV.\cite{gammaenergies} The compound formation energy is given by the Gamow window. \cite{energywindows} For neutron captures, the upper end of the relevant energy window is at most at 0.2 MeV, even at the high plasma temperatures encountered in explosive burning environments. This translates into an excitation energy of the compound nucleus of $S_\mathrm{n}+0.2$ MeV (with the neutron separation energy $S_\mathrm{n}$). The energy window of charged particles is shifted to slightly higher energy due to the Coulomb barrier but does not exceed several MeV.\cite{raureview,energywindows} Measuring at much larger energy than the astrophysical one does not yield much relevant information in most cases. This is because at higher energy the cross sections show a different sensitivity (see Sec.\ \ref{sec:sensi}) to photon- and particle-strengths than at astrophysical energies and higher partial waves also play a role.\cite{raureview,sensi} Furthermore, additional reaction mechanisms may occur which are not appearing in the astrophysical energy range.\cite{raureview} At high energy, so many additional effects are coming into play that it becomes difficult to impossible to extract and compare the ones relevant in the astrophysical energy range.

Much can be learned from photodisintegration experiments (subsuming experiments with real photons and Coulomb dissociation). However, it should be clear that there are severe limitations in these methods regarding astrophysical rates for heavier nuclei, despite their success in light nuclear systems. Without further information (such as ($\gamma$,$\gamma'$) data), a ($\gamma$,n) measurement cannot be used to constrain the astrophysical rate or to even only test the reliability of model predictions of such rates. Data showing the relative particle emission to g.s.\ and excited states, though, may test to a certain extent the prediction of thermal modification of the stellar capture reaction and thus of $X_\mathrm{exc}$. Also the energy-dependence of optical potentials can be studied at low energy because of the decreasing relative transition energies for reactions leading to excited states in the final nucleus. The situation is similar in photodisintegrations emitting charged particles. If measurements could study, however, charged particle emission below the Coulomb barrier, this would constrain the rate because at low relative energy the cross section is determined by the charged particle width (as can be verified by looking at the sensitivities $s$).\cite{sensi} Unfortunately, such cross sections are very small and outside the reach of current methods, especially for unstable nuclei.

The above applies also to indirect studies of reactions in other nucleosynthesis processes, such as described in Secs.\ \ref{sec:sproc}$-$\ref{sec:nup}.

\subsection{Proton-rich burning off stability}
\label{sec:nup}

The fact that the proton dripline is closer to the line of stability than the neutron dripline may not be taken as an indication that fewer problems are encountered on the proton-rich side. In general, the effective energies contributing to the reaction rate integral in Eq.\ (\ref{eq:stellrate}) are higher for charged particle reactions than when dealing with neutron captures on neutron-rich nuclei.\cite{energywindows} Moreover, temperatures in nucleosynthesis processes on the proton-rich side are higher than on the neutron rich side, typically $T>1$ GK, because otherwise the rates would be too slow for short, explosive timescales. This somewhat alleviates the problem of the reaction mechanisms as discussed in Sec.\ \ref{sec:nld}. The Hauser-Feshbach model seems valid up to close to the proton dripline, even for proton captures.\cite{raureview,rtk} As pointed out before, this is a challenge for experimental studies using indirect methods (except at the proton dripline) but eases a theoretical treatment.

Nucleosynthesis beyond Fe on the proton-rich side not only requires higher temperatures but also a high proton number density. An important process at such conditions is the rapid proton capture process or $rp$ process. A certain type of regular X-ray bursts (type I) is explained by an $rp$ process in proton-rich matter accreted on the surface of a neutron star in a binary system.\cite{schatz,endpoint,wooxray} Nuclear burning on the surface of neutron stars is discussed elsewhere in this volume.\cite{schatzthis} Therefore I focus on another proton-rich process in the following, the neutrino-induced (or neutrino-aided) rapid proton capture or $\nu p$ process.

Hydrodynamical simulations indicate that part of the innermost, still ejected layers of ccSN may become proton-rich due to the interaction with the intense $\nu_\mathrm{e}$ wind (and $\overline{\nu}_\mathrm{e}$ wind) within which they move outward. The matter quickly cools from the initially high temperature, assembling nucleons mainly to $^{56}$Ni and $\alpha$-particles in a nuclear statistical equilibrium, leaving a large number of free protons. At sufficiently low temperature ($\leq 3$ GK), rapid proton captures ensue on $^{56}$Ni. Production of heavier nuclei beyond $^{64}$Ge would not be possible without a tiny fraction of free neutrons which is created by $\overline{\nu}_\mathrm{e}$ captures on the free protons. The supply of free neutrons in the neutrino-wind allows (n,p) reactions bypassing slow electron captures and $\beta^+$ decays. This provides the possibility that sequences of proton captures and (n,p) reactions produce nuclei with larger and larger $Z$ and $A$.\cite{fro06a,fro06b,pruet06} Charged particle reactions then freeze out quickly below 1.5 GK, leaving only (n,p) and (n,$\gamma$) acting at late time which push the matter back to stability.

\begin{figure}
\begin{center}
  \includegraphics[width=1.1\textwidth]{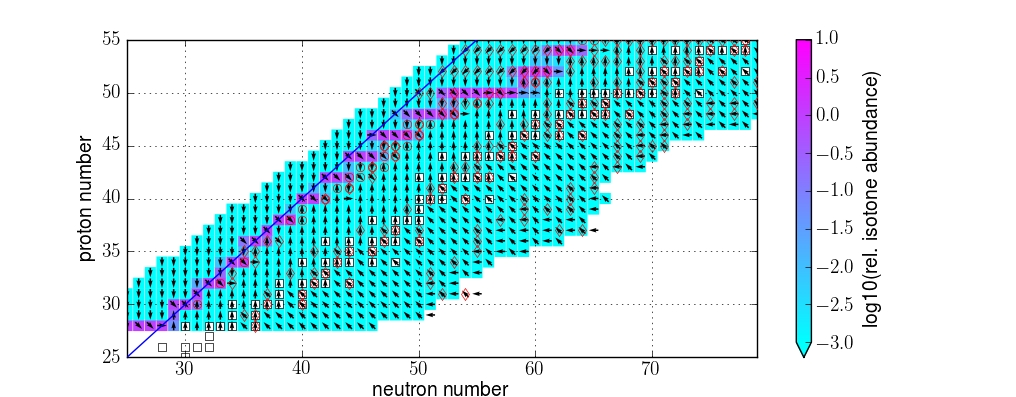}
  \caption{\label{fig:equiabuns}The $\nu p$ process: Equilibrium abundance in each isotonic chain (color-shaded), dominating reactions (arrows), and nuclear uncertainties (circles and diamonds). Stable nuclei are marked as white squares and the $N=Z$ line is shown.}
\end{center}
\end{figure}

\begin{figure}
\begin{center}
  \includegraphics[width=1.1\textwidth]{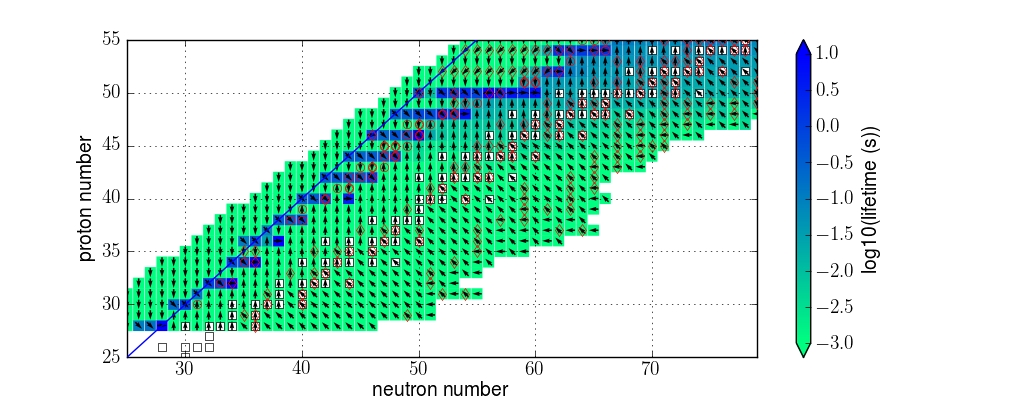}
  \caption{\label{fig:flux}Same as Fig.\ \ref{fig:equiabuns} but the color shade of each nucleus gives its lifetime against the dominating destruction reaction.}
\end{center}
\end{figure}

The location of the effective $\nu$p-process path remains remarkably unaffected by variations of the astrophysical parameters, such as entropy, expansion timescale, and details of the reverse shock \cite{frorautru,raufro,fro06b,pruet06,wanajo11}. Also systematic variations of reaction rates show only small effects, if any, regarding the path location \cite{frorautru,raufro,wanajo11}. All these variations, however, determine how far up the path is followed or whether it is terminated already at low $Z$. Thus, also the achieved abundances within the path are determined by these conditions. The reason for this behavior can easily be understood when realizing that (p,$\gamma$)-($\gamma$,p) equilibrium is upheld until late times. In consequence, the abundance maximum in each isotonic chain is given by the equilibrium abundance (Eq.\ \ref{eq:equilcap}), color-shaded for each nucleus in Fig.\ \ref{fig:equiabuns}. The corresponding Fig.\ \ref{fig:flux} shows the same rate field but with the lifetimes of the nuclei. These
\emph{rate field plots} are extremely helpful to examine a nucleosynthesis process. The shade of each nucleus in the plot is either its relative abundance within an isotopic chain (Fig.\ \ref{fig:equiabuns}) or related to its lifetime with respect to the fastest reaction destroying it. The arrows give the direction of the fastest \emph{net rate per target nucleus} $r^*/n_A$ (see also Eq.\ \ref{eq:stellrategeneral}). The net rate is obtained from the difference of a rate per target nucleus and its reverse rate. These are not the actual rates in a reaction network as these would also depend on the abundance of the target nucleus but they allow to quickly gather which reaction dominates on a nucleus and which direction the synthesis path would take, were the nucleus actually produced.

\begin{figure}
\begin{center}
\includegraphics[width=1.1\textwidth]{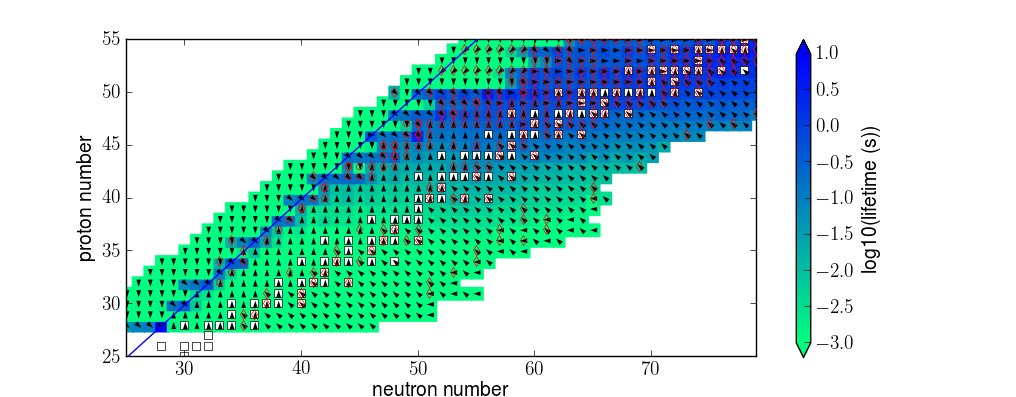}
\caption{\label{fig:lateflux}Same as Fig.\ \ref{fig:flux} but at very late times; neutron captures drive the path even further to stability.}
\end{center}
\end{figure}

Due to the (p,$\gamma$)-($\gamma$,p) equilibrium each proton-rich nucleus in an isotonic chain is rapidly (compared to the expansion timescale) converted into the nucleus favored by the equilibrium conditions. Therefore, the maximum within a chain is characterized by a low (p,$\gamma$) reaction $Q$ value because the relative speed of forward and reverse rate depends exponentially on the $Q$ value (Eq.\ \ref{eq:revphoto}). The largest flux into the next isotonic chain occurs at these nuclei, which are also called waiting points. As can be seen in Fig.\ \ref{fig:flux}, these nuclei indeed exhibit longer lifetimes but they may still be overcome if the neutron abundance is sufficiently high. This implies that a variation of the neutron density or the (n,p) rate on these waiting points will mostly affect how fast nuclei with larger $Z$ can be reached within the timescale given by the expansion.

Another nuclear-structure determined feature can be seen immediately in Figs.\ \ref{fig:equiabuns} and \ref{fig:flux}: while the waiting points follow the $Z=N$ line up to Mo, they are extending further and further to neutron-richer isotopes between Mo and Sn, pushing the path gradually away from the $Z=N$ line. The path is pushed strongly towards stability at the Sn isotopes and above, providing a strong barrier for the efficient production of any elements beyond Sn. Two effects are acting here: The location of the waiting-points at neutron-richer nuclides with longer lifetimes with respect to both decays and (n,p) reactions, and the fact that also the proton captures become slower due to the higher Coulomb barriers at larger $Z$. The latter leads to a dominance of (n,$\gamma$) reactions over (p,$\gamma$) and ($\gamma$,p) as can be seen in Figs.\ \ref{fig:equiabuns} and \ref{fig:flux} at high $Z$, and especially in Fig.\ \ref{fig:lateflux} showing the situation at late times. The neutron captures push the nucleosynthesis strongly towards stability and even beyond and prevent the production of appreciable amounts of matter above the Sn region.

The above general conclusions hold for any trajectory yielding an appreciable $\nu$p-process. Moreover, the range of conditions permitting such a process are also set by nuclear properties. Without neutrons, the waiting points cannot not be passed but a large neutron density (caused, e.g., by a higher $\overline{\nu}_\mathrm{e}$ flux) would make the (n,p) reactions faster than (p,$\gamma$) on \emph{any} proton-rich nuclide above $^{56}$Ni. On the other hand, too high a temperature prevents the outbreak from the $^{56}$Ni seed as ($\gamma$,p) and ($\gamma$,$\alpha$) reactions are faster than proton captures and always establish QSE with $^{56}$Ni. Finally, in a freeze-out charged-particle captures are suppressed according to the Coulomb barriers and a further increase in $Z$ is more and more hampered. The exact values determining the window of favorable conditions can thus be estimated from nuclear properties in the spirit of the classical B$^2$FH paper.\cite{b2fh}
The astrophysical hydrodynamics just determines how long the ejected matter is subjected to those favorable conditions and thus how far up in $Z$ and $A$ $\nu$p nucleosynthesis can proceed.

The uncertainties in the astrophysical conditions are similar to those for the $r$ process in ccSN (Sec.\ \ref{sec:rproc}) because they stem from the same problems in the supernova modelling. The explosion mechanism and the neutrino interactions not only determine how much matter is ejected at high entropy from the innermost layers and the ejection speed but also influence the time evolution of the neutron-to-proton ratio. The details of the hydrodynamics also determine the wind termination when the supersonic inner layers run into the slower outer layers of the exploding star and abruptly decelerate.\cite{wanajo11}

Another possible problem concerning the $\nu_\mathrm{e}$ and $\overline{\nu}_\mathrm{e}$ spectra close above the neutrino spheres was pointed out recently.\cite{marti,roberts1,roberts2} The difference in mean field potentials between neutrons and protons was incompletely considered in previous studies. Including an improved treatment seems to render the ejected layers less proton-rich. The actual impact on the $\nu p$ process is pending consistent nucleosynthesis studies. Even when using the improved mean field effects, however, the remaining hydrodynamical uncertainties may be too large to arrive at a definitive conclusion regarding the significance of the $\nu p$ process.

On the nuclear reaction side, the results are mostly insensitive to proton captures due to the prevailing (p,$\gamma$)-($\gamma$,p) equilibrium. Only at late freeze-out times, this equilibrium is left and gives rise to some sensitivity to a variation of rates.\cite{frorau,frorautru} The equilibrium abundances, in turn, are determined by the proton separation energies $S_\mathrm{p}$ and thus by nuclear masses.
In Figs.\ \ref{fig:equiabuns}--\ref{fig:lateflux} circles mark nuclei where remaining experimental uncertainties in the $Q$ values (mass differences) affect the $\nu$p-process results. It should be noted that the waiting points at lowest $S_\mathrm{p}$ within an isotonic chain are somewhat different from waiting points in the $rp$ process. The proton number density is much lower in the $\nu p$ process than in the $rp$ process and therefore there are no cases of uncertain waiting points which may be overcome by proton captures when the masses are changed within uncertainties. Therefore also $^{64}$Ge still is a waiting point, although it is not anymore in the $rp$ process after a recent mass measurement.\cite{tu} Changing the masses of the nuclei marked in the figures, nevertheless, affects the abundance pattern along the $\nu p$ path which is obtained after decay to stability and may also impact up to which mass significant abundances can be created. Another difference to $rp$ waiting points is the fact that (n,p) reactions are much faster than $\beta$-decays and electron captures and therefore there is no actual ``waiting'' in the reaction flow, contrary to $rp$- and $r$ process waiting points.

The flow to heavier nuclei is determined by (n,p) reactions on the waiting point nuclei (Figs.\ \ref{fig:equiabuns}--\ref{fig:lateflux}) and thus a knowledge of these is essential.\cite{wanajo11,frorau,frorautru} Neutron captures on proton-rich nuclei are of some relevance at large $Z$ and/or at late times (Fig.\ \ref{fig:lateflux}). Because of the large $S_\mathrm{n}$ of proton-rich, unstable nuclei, (n,p) reaction $Q$ values are large and thus $E_\mathrm{form}$ is large and the Hauser-Feshbach model will be applicable. As pointed out in Sec.\ \ref{sec:nld}, this is bad news for experiments probing direct transitions or relying on the fact that only few transitions contribute. On the other hand, even though the Hauser-Feshbach model is well established, its input has to be determined reliably, especially for the cases where it predicts competitions between rates. In Figs.\ \ref{fig:equiabuns}--\ref{fig:lateflux} diamonds mark targets for which (p,$\gamma$), (n,p), or (n,$\gamma$) rates are close and theoretical uncertainties in the predicted rates may be important. These occur only at large $Z$, reached at late time if at all.

A special class of reactions are those which govern the onset of the $\nu p$-process at high temperature. Since the details depend on competitions between reactions, they are mostly trajectory-independent and it is possible to identify a set of reactions for which it would be desireable to know the rates (or ratios of rates of possibly competing reaction types) with good accuracy.

When freezing out from nuclear statistical equilibrium at high temperature, the $\nu p$ process is delayed by several issues. At high temperature, ($\gamma$,p) reactions are fast and the equilibrium abundances are always located around $^{56}$Ni. Since the main abundance is concentrated in $^{56}$Ni, further processing is halted until the $^{56}$Ni waiting point can be bridged effectively and the (p,$\gamma$)-($\gamma$,p) equilibrium abundance maxima in the subsequent isotonic chains are moved to higher $Z$.\cite{frorautru,wanajo11} This depends on the competing rates of ($\gamma$,p), (n,$\gamma$), and (n,p) on $^{56}$Ni and occurs at $T\approx 3.5$ GK.

\begin{figure}
\begin{center}
\includegraphics[width=1.1\columnwidth]{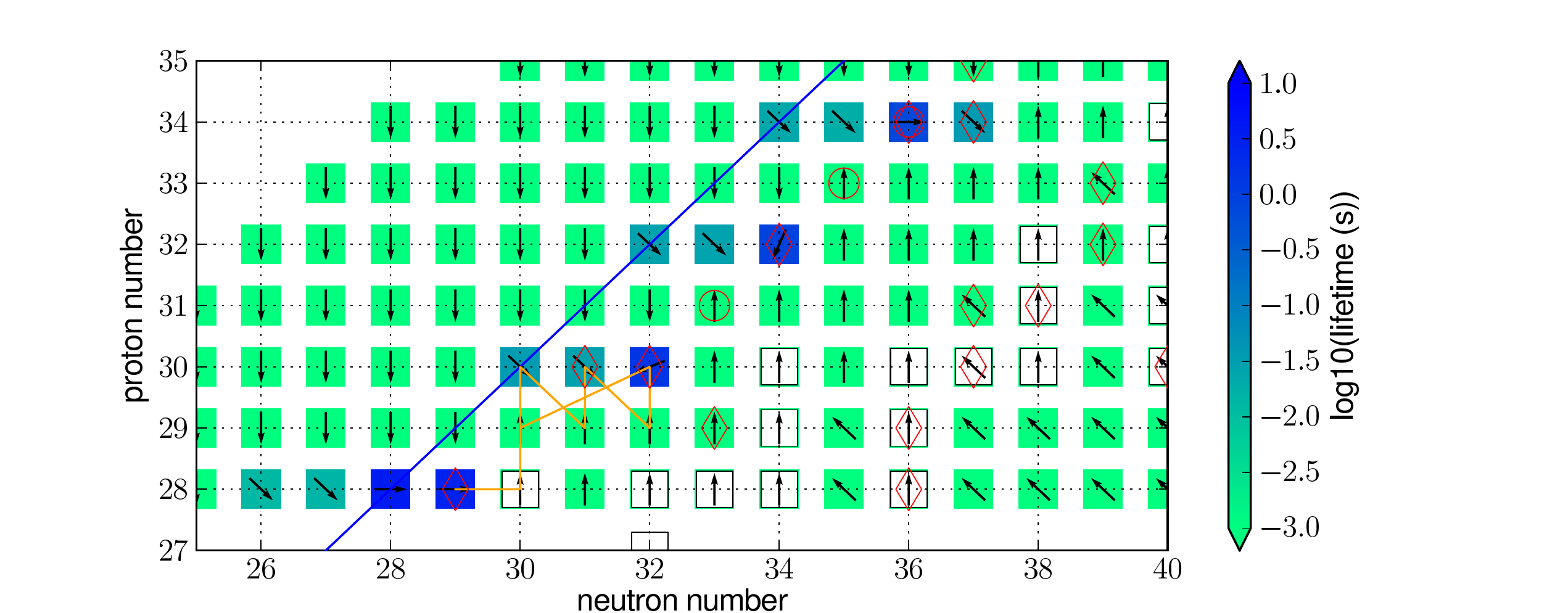}
\caption{\label{fig:breakout1}Rate field plot showing an example of a possible limiting cycle at Zn in the $\nu p$ process, the main path is drawn to guide the eye.}
\end{center}
\end{figure}

\begin{figure}
\begin{center}
\includegraphics[width=1.1\columnwidth]{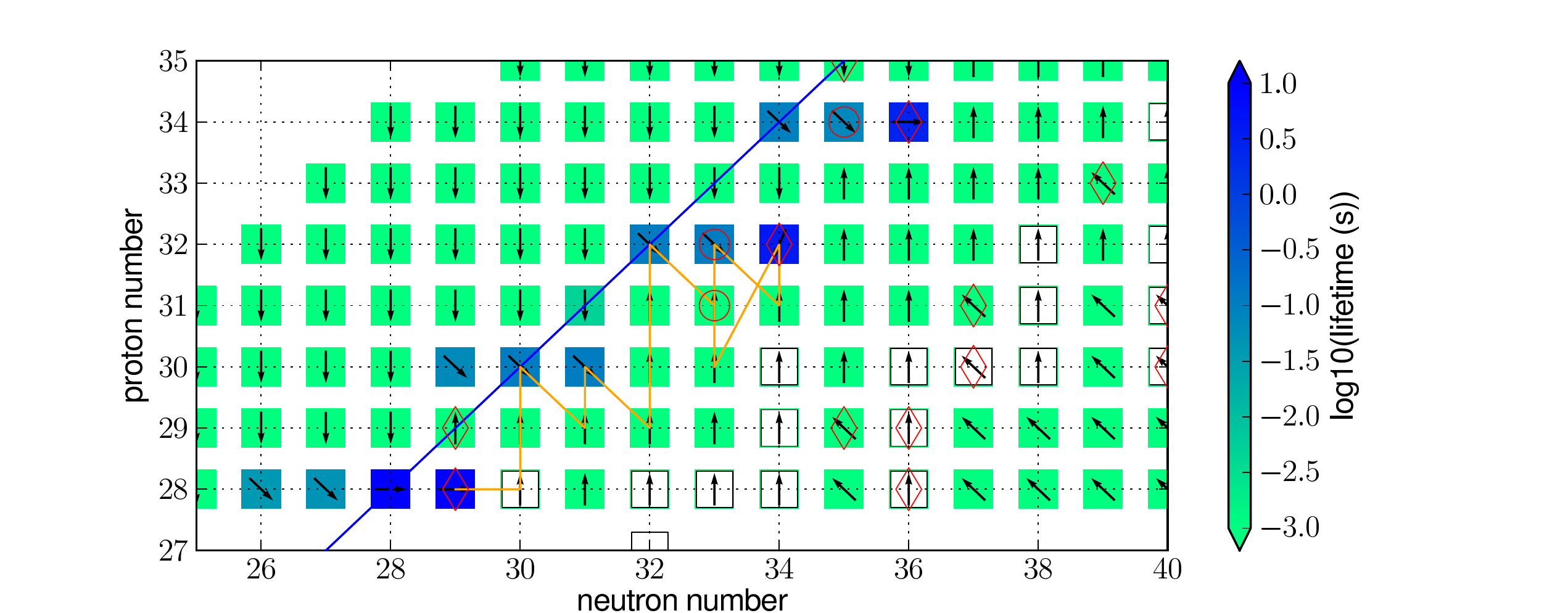}
\caption{\label{fig:breakout2}Rate field plot showing an example of a possible limiting cycle at Ge in the $\nu p$ process, the main path is drawn to guide the eye.}
\end{center}
\end{figure}

Whether further processing occurs already at this temperature, however, depends on the relative speeds of ($\gamma$,$\alpha$), (p,$\alpha$), and (n,$\alpha$) reactions on waiting point isotopes of Zn and Ge compared to the (n,p), (n,$\gamma$), or (p,$\gamma$) reactions required to commence with nucleosynthesis to heavier elements. We find many of these rates to be comparable within the theoretical uncertainties and therefore they have to be considered as important competition points at early times.

To illustrate the importance of accurately knowing the relative strengths of these reactions, two examples are presented here. The reactions  $^{62}$Zn($\gamma$,$\alpha$)$^{58}$Ni, $^{62}$Zn(p,$\alpha$)$^{59}$Cu, and $^{66}$Ge(n,$\alpha$)$^{63}$Zn are close to the competing (n,p), (n,$\gamma$), or (p,$\gamma$) reactions. The flow beyond Ge would be considerably impeded were these reactions dominating because the reaction cycles shown in Figs.\ \ref{fig:breakout1} and \ref{fig:breakout2} would form. Initially the first cycle would be due to the $^{62}$Zn($\gamma$,$\alpha$)$^{58}$Ni reaction being faster than other reactions on $^{62}$Zn, hindering any production beyond Zn. While cooling down, the situation would change to the one shown in Fig.\ \ref{fig:breakout1} where $^{62}$Zn($\gamma$,$\alpha$) has become slow but there still would be a cycle caused by the reaction $^{62}$Zn(p,$\alpha$)$^{59}$Cu. To know whether this cycle occurs, the relative strengths of $^{62}$Zn(p,$\alpha$), $^{62}$Zn(n,$\gamma$), $^{62}$Zn(n,p), $^{62}$Zn(n,$\alpha$) have to be known. In the current rate set, $^{62}$Zn(p,$\alpha$), $^{62}$Zn(n,$\gamma$), and $^{62}$Zn(n,$\alpha$) are quite comparable down to $T\approx 2.3$\,GK.

Since a (p,$\alpha$) reaction is competing with neutron-induced reactions in this possible Zn cycle, one might wonder whether an increase in $\rho_\mathrm{n}$ would also help to avoid the cycle. This can only be achieved by a larger $Y_\mathrm{n}$ because a simple scaling of the density $\rho$ equally affects the (p,$\alpha$) reaction. There are two problems with this, however. Firstly, both (n,$\gamma$) and (n,$\alpha$) would be increased and the competiton between these reactions within their uncertainties would not be lifted. Secondly, to make the neutron-induced reactions faster than (p,$\alpha$), the $Y_\mathrm{n}$ has to be increased so much that (n,$\alpha$) or (n,p) reactions become faster than (p,$\gamma$) for all nuclei in the subsequent isotonic chains, thereby dissolving the (p,$\gamma$)-($\gamma$,p) equilibrium and pushing nucleosynthesis towards stability without the possibility to further move up in $Z$. Nevertheless, if $Y_\mathrm{n}$ were increased moderately, this avoided a dominating (p,$\alpha$) reaction but required some fine-tuning.

In the second example, a cycle would form at Ge if the reaction $^{66}$Ge(n,$\alpha$)$^{63}$Zn was dominating, as shown in Fig.\ \ref{fig:breakout2}. Using the current rate set, the reaction $^{66}$Ge(n,$\alpha$)$^{63}$Zn is found to be in strong competition with $^{66}$Ge(n,$\gamma$) but a re-evaluation of the reaction rates may change this picture. This cycle would be robust with respect to changes in $Y_\mathrm{n}$.

Another cycle, close to $^{56}$Ni and acting down to $T\approx 3$ GK was previously found:\cite{arcofromart} The reaction $^{59}$Cu(p,$\alpha$)$^{56}$Ni cycles matter back to $^{56}$Ni. This cycle is also affected by possible competitions with other reactions. This nicely illustrates the uncertainties in $\nu p$ nucleosynthesis due to reaction rates and underlines the importance of studying competing reactions. To better constrain the efficiency of the $\nu p$ process from the nuclear physics side, it is necessary to establish tight bounds on the relative strengths of the strongly competing (p,$\gamma$), (p,$\alpha$), (n,$\gamma$), (n,p), and (n,$\alpha$) reactions on targets from Ni to Ge and with $Z\leq N\leq Z+3$.

These conclusions concerning the important rate competitions do not strongly depend on the chosen trajectory. The ratio of proton- to neutron-induced rates is independent of density changes. Modifying the density at a given temperature affects the ratio of particle-induced to photodisintegration reactions and thus leads to altered break-out temperatures from $^{56}$Ni or from (p,$\gamma$)-($\gamma$,p) equilibrium. Since the temperature dependence of the reactivities, however, is exponential and the density dependence only linear, the break-out temperatures are shifted only slightly. The same is true when varying $Y_\mathrm{p}$ alone. Due to the densities encountered in the ejected layers, $n_\mathrm{p}$ always stays well below the values of an $rp$ process, where (p,$\gamma$) would dominate. Moreover, a realistic range of entropies in ejected proton-rich layers implies only a narrow range of temperatures at a given density. The only strong astrophysical dependence is on $Y_\mathrm{n}$ created by the $\overline{\nu}_\mathrm{e}$ flux present at a given temperature. However, this does not change the ratio between (n,$\gamma$), (n,p), and (n,$\alpha$) reactions, the latter being a hindrance to the flow up to heavier nuclei. Furthermore, as pointed out above, there is not much room for modification of $Y_\mathrm{n}$ without destroying the possibility for $\nu p$ nucleosynthesis. A low $Y_\mathrm{n}$ prevents $\nu p$ processing because of slow (n,p) and (n,$\gamma$) reactions but a much higher $Y_\mathrm{n}$ also does not help the break-out as discussed above.

Nevertheless, although the rate competitions are robust with respect to changes in the hydrodynamic evolution, the impact of, for example, cycles as discussed above depends on the trajectory. The time evolution of the conditions within a trajectory determines how long the favorable conditions for a cycle (if existing) are upheld. If the relevant temperature-density-$Y_\mathrm{n}$ regime for a possible cycle is crossed quickly whereas the later evolution proceeds slower, leaving enough time for further $\nu p$ processing, then the relevance of the cycle is diminished. For example, although the cycle formed by the $^{59}$Cu(p,$\alpha$)$^{56}$Ni reaction is appearing at a given temperature, it has only moderate impact on the final abundances when chosing a regular freeze-out without significant wind termination shock. Using a trajectory in which the wind termination shock keeps the temperature at slightly above 3 GK for an extended time will yield a large impact, however.\cite{arcofromart}

In conclusion, although the astrophysical constraints seem to be similarly weak for the $\nu p$ process as for the $r$ process, it is better constrained by nuclear physics and exhibits smaller uncertainties therein, at least in the dominating rates. Nevertheless, an experimental verification of the predicted rates will be difficult, not only because of the short-lived, intermediate and heavy nuclei involved but also due to the high plasma temperatures, giving rise to considerable thermal excitation and thus small g.s.\ contributions $X_0$ to the stellar rate.

\section{Summary}
\label{sec:conclusion}

Nucleosynthesis beyond Fe poses additional challenges not encountered when studying astrophysical processes involving light nuclei. It requires different approaches, both in theory and experiment. The main considerations were presented for a selection of nucleosynthesis processes and reactions. The presentation should not be viewed as confining the discussed problems to the specific processes. The intention was to generally introduce the concepts and possible pitfalls along with some examples. Similar problems may apply to further astrophysical processes involving nuclei from the Fe region upward and/or at high plasma temperatures. The framework and strategies presented here shall aid the conception of experimental and theoretical approaches to further improve our understanding of the origin of trans-iron nuclei.

\section*{Acknowledgments}
This work was supported by the Swiss NSF, the European Research Council, and the THEXO collaboration within the 7$^{th}$ Framework Program ENSAR of the EU.

\end{document}